\documentclass[aps,prd,preprintnumbers,showpacs,twocolumn,groupedaddress,nofootinbib]{revtex4}
\usepackage{graphicx}
\usepackage{latexsym}
\def\beq{\begin{equation}}
\def\eeq{\end{equation}}
\def\bey{\begin{eqnarray}}
\def\eey{\end{eqnarray}}

\def\lsim{\mathrel{\raise.3ex\hbox{$<$\kern-.75em\lower1ex\hbox{$\sim$}}}}
\def\gsim{\mathrel{\raise.3ex\hbox{$>$\kern-.75em\lower1ex\hbox{$\sim$}}}}

\begin{document}

\title{A Leptophobic $Z'$ And Dark Matter From Grand Unification}  
\author{Matthew R.~Buckley$^1$, Dan Hooper$^{1,2}$, and Jonathan L.~Rosner$^3$}
\affiliation{$^1$Center for Particle Astrophysics, Fermi National Accelerator Laboratory, Batavia, IL 60510, USA}
\affiliation{$^2$Department of Astronomy and Astrophysics, University of Chicago, Chicago, IL 60637, USA}
\affiliation{$^3$Department of Physics, University of Chicago, Chicago, IL 60637, USA}

\date{\today}

\begin{abstract}

We explore the phenomenology of Grand Unified Models based on the $E_6$ group, focusing on the $Z'$ with suppressed couplings to leptons that can appear in such models. We find that this $Z'$ can accommodate the $W$+dijets anomaly reported by the CDF collaboration. Furthermore, a viable dark matter candidate in the form of a right-handed sneutrino is also present within the fundamental 27-dimensional representation of $E_6$. Through its sizable couplings to the $Z'$, the dark matter is predicted to possess an elastic scattering cross section with neutrons which can generate the signals reported by the CoGeNT and DAMA/LIBRA collaborations. To avoid being overproduced in the early universe, the dark matter must annihilate to leptons through the exchange of charged or neutral fermions which appear in the {\bf 27} of $E_6$, providing an excellent fit to the gamma ray spectrum observed from the Galactic Center by the Fermi Gamma Ray Space Telescope.

\end{abstract}

\pacs{12.10.Dm,14.70.Pw,14.80.-j,95.35.+d; FERMILAB-PUB-11-278-A, EFI 11-15}
\maketitle

\section{A Leptophobic $Z'$ Boson From Grand Unification}

The CDF collaboration has recently announced the observation of an excess of events containing a lepton (electron or muon), missing transverse energy, and two jets~\cite{cdf}. This can be interpreted as the production of a new state with a mass of $\sim$150-160 GeV which decays to quarks or gluons, along with a Standard Model $W^{\pm}$ decaying to an electron or muon along with a neutrino.  This excess of events, which was observed in 7.3 fb$^{-1}$ of CDF data, has been reported to be statistically significant at a level of 4.1$\sigma$. Although the D0 collaboration does not find evidence of such a sizable excess in their analysis of 4.3 fb$^{-1}$ of data~\cite{D0}, a signal roughly half as large as reported by CDF might be consistent with both experimental results.

Although several explanations have been proposed for the CDF excess~\cite{Buckley:2011vc,Yu:2011cw,Wang:2011uq,Eichten:2011sh}, perhaps the simplest is the introduction of a new gauge boson~\cite{Buckley:2011vc,Yu:2011cw,Wang:2011uq}. In order to evade constraints from LEP II and the Tevatron, however, such a state must have couplings to leptons (or more specifically, to electrons and muons) which are suppressed by at least a factor of 4-5 relative to those to quarks~\cite{Buckley:2011vc,Yu:2011cw}. A gauge boson with these characteristics is commonly known as a leptophobic $Z'$.

New $U(1)$ gauge groups and the $Z'$ bosons which accompany them are found in many phenomenologically viable and well motivated extensions of the Standard Model~\cite{Langacker:2008yv}. Such extensions introduce anomalies, however, whose cancellation requires the addition of new fermions. The sole exception to this rule is a $Z'$ which couples to baryon minus lepton number. Thus any anomaly-free, flavor universal model that contains a leptophobic $Z'$ must invariably contain new fermions.

Additional $U(1)$ gauge groups are a feature common to many grand unified theories (GUTs), including those based on the groups $SO(10)$ and $E_6$  (the $SU(5)$ group is of the same rank as the Standard Model, and thus does not contain any additional $U(1)$'s). The $Z'$ which arises in $SO(10)$ models, however, is not leptophobic. As the $E_6$ group is rank 6 (two higher than the gauge group of the Standard Model), GUTs based on this group can include two additional $U(1)$'s, leading to a variety of possible charge assignments and interactions~\cite{London:1986dk}. For this reason, and as $E_6$ is well motivated  within the context of $E_8 \times E_8$ string theory~\cite{string}, we will focus on supersymmetric GUTs built upon this gauge group.\footnote{Non-supersymmetric $E_6$ models are of course possible. However, for this initial study, we choose this more constrained scenario.} For a review of $E_6$ models or of GUTs in general,  see Refs.~\cite{Hewett:1988xc} and~\cite{Langacker:1980js}, respectively. 

$E_6$ can be broken such that $E_6 \rightarrow SO(10) \times U(1)_{\psi} \rightarrow SU(5) \times U(1)_{\chi} \times U(1)_{\psi}$, of which the $SU(5)$ contains the gauge group of the Standard Model. The charges of any $U(1)'$ and the corresponding $Z'$ boson that appears as part of the low-energy theory can be written as a linear combination of the charges of the two additional $U(1)$'s in $E_6$:
\begin{equation}
Q(\alpha) =  \cos \alpha\, Q_{\chi} + \sin \alpha\, Q_{\psi}.
\end{equation}
Although for no choice of $\alpha$ is $Q(\alpha)$ completely leptophobic, for the well motivated value of $\alpha=-\tan^{-1} \sqrt{5/3}$ (known as the $\eta$-model of $E_6$~\cite{etamodel}, such that $Q_{\eta} \equiv \sqrt{3/8}\, Q_{\chi}+\sqrt{5/8}\, Q_{\phi}$), kinetic mixing can lead to a $Z'$ with highly suppressed couplings to leptons~\cite{E6,E62,Barger:1996kr}. In particular, although the $U(1)_Y$ and $U(1)_{\eta}$ gauge groups undergo no kinetic mixing at the $E_6$ breaking scale, if the effective particle content at some lower energy scale does not contain full representations of $E_6$, kinetic mixing will be generated by renormalization group (RG) evolution~\cite{E6,Rizzo:1998ut}. The full Higgs sector is expected to be split in mass, as some of the scalar fields will have GUT-scale vacuum expectation values (vevs). Thus, during the RG running down to the weak scale, the particle content is such that a non-zero mixing parameter will evolve.

\begin{table*}
  \centering
  \begin{tabular}{l@{\quad}c@{\qquad}c@{\qquad}c@{\qquad}c@{\qquad}c@{\qquad}c@{\qquad}c@{\qquad}c@{\qquad}c@{\qquad}c}
          & \vline &Color &$Q_{\rm EM}$& $\sqrt{5/3}\, Y$& $2\sqrt{6} \,Q_{\psi}$ & $2\sqrt{10}\,Q_{\chi}$&$2\sqrt{15}\, Q_{\eta}$ &\vline&$2\sqrt{3/5}\,Q'\,(\delta=1/3)$ &    \\\hline
    $Q=\bigg(\begin{array}{c}u\\d\end{array}\bigg)_L$   & \vline & ${\bf 3}$&  $\bigg(\begin{array}{c}2/3\\$-1/3$\end{array}\bigg)$  & 1/6  &1  & -1 & -2 & \vline& -1/3  &  \\ 
    $\bar{u}_{\raisebox{-1pt}{\tiny \it R}}$ & \vline &${\bf \bar{3}}$&-2/3& -2/3 &1  & -1 & -2 & \vline& -2/3  &  \\ 
\vspace{-0.3cm} 
     & \vline &&& &  &  &  & \vline&   &   \\ 
  $\bar{d}_{\raisebox{-1pt}{\tiny \it R}}$ & \vline &${\bf \bar{3}}$&1/3& 1/3  &1  &  3 &  1 & \vline&  1/3  &  \\
    $L=\bigg(\begin{array}{c}\nu\\e\end{array}\bigg)_L$ & \vline &${\bf 1}$&  $\bigg(\begin{array}{c}0\\$-1$\end{array}\bigg)$    & -1/2 &1  &  3 &  1 &
 \vline&   0   &  \\
    $e^+_{\raisebox{-1pt}{\tiny \it R}}$ & \vline &${\bf 1}$&1& 1    &1  & -1 & -2 & \vline&   0   &  \\
\vspace{-0.3cm} 
     & \vline &&& &  &  &  & \vline&  &\\ \hline
    $\bar{H}=\bigg(\begin{array}{c}\bar{E}\\\bar{N}\end{array}\bigg)_R$ & \vline &${\bf 1}$&  $\bigg(\begin{array}{c}1\\0\end{array}\bigg)$    & 1/2  &-2  &   2 &  4 & \vline&  1  &  \\
    $H=\bigg(\begin{array}{c}N\\E\end{array}\bigg)_L$ & \vline &${\bf 1}$& $\bigg(\begin{array}{c}0\\$-1$\end{array}\bigg)$& -1/2  &-2  & -2 &  1 & \vline&  0  &  \\
    $h_{\raisebox{-1pt}{\tiny \it L}}$ & \vline &${\bf 3}$&-1/3& -1/3  &-2  &  2 &  4 & \vline&   2/3  &  \\
\vspace{-0.3cm} 
     & \vline &&& &  &  &  & \vline&   &  \\ 
    $\bar{h}_{\raisebox{-1pt}{\tiny \it R}}$ & \vline &${\bf \bar{3}}$&1/3& 1/3  &-2  & -2 &  1 & \vline&  1/3  \\
\vspace{-0.3cm} 
     & \vline &&& &  &  &  & \vline&   &  \\ 
  $\bar{\nu}_{\raisebox{-1pt}{\tiny \it R}}$ & \vline &${\bf 1}$&0& 0  &1  &  -5 &  -5 & \vline&  -1  \\
\vspace{-0.3cm} 
     & \vline &&& &  &  &  & \vline&   &  \\ 
    $S_{\raisebox{-1pt}{\tiny \it L}}$ & \vline &${\bf 1}$&0& 0      & 4  &  0 &  -5 & \vline&  -1  \\ \hline
  \end{tabular}
  \caption{Properties of the fermions in a {\bf 27} of $E_6$. For a value of $\delta=1/3$, the $Z'$ is completely leptophobic. In addition to its couplings to Standard Model quarks, the $Z'$ is predicted to have large couplings to the dark matter candidates in this model (the superpartners of $\bar{\nu}_R$ and $S_L$).}
  \label{tab:summary}
\end{table*}

%%%

%%%

After accounting for the effects of kinetic mixing, the effective charges of the fermions to the $Z'$ are given by $Q'(\delta)=Q_{\eta}+\delta Y$, where the values of $Q_{\psi}$, $Q_{\chi}$ and $Q_{\eta}$, are given in Table I for all of the fermions in the fundamental 27-dimensional representation of $E_6$. For the choice of $\delta=1/3$, the $Z'$ is completely leptophobic ($Q'_L=Q'_{e^+}=0$).  The size of $\delta$ is highly dependent on the composition and mass scale of the Higgs sector; while the minimal fields required to give masses to the matter {\bf 27}s are not sufficient to generate $\delta \approx 1/3$, we would not expect these to be the only fields of relevance all the way up to the unification scale. In Ref.~\cite{E6}, for example, a reasonable Higgs sector was described which naturally leads to $\delta\approx 1/3$. For purposes of explaining the CDF $W+$dijets excess, we do not necessarily require complete leptophobia, and values in the range of $\delta\approx0.25-0.40$ are acceptable.\footnote{For another way to introduce a leptophobic $Z'$, one may also consider models in which baryon number and lepton number are gauged~\cite{darkforces,Barger:1996kr}. Alternatively, a $Z'$ could acquire couplings to Standard Model quarks through higher dimensional effective operators~\cite{effective}.} Throughout the remainder of this paper, we carry out our calculations using the choice of $\delta=1/3$.

With mixing between $Y$ and $\eta$, the photon, $Z$ and $Z'$ will be linear combinations of the neutral gauge bosons of the $U(1)_Y$, $U(1)_\eta$, and $SU(2)_L$  gauge groups. This induces corrections to the electroweak precision observables, which place limits on leptophobic $E_6$ models from LEP and other precision measurements. Although constrained, some regions of mixing parameter space remain allowed for a leptophobic $Z'$ of mass $\sim$150 GeV~\cite{E6,Umeda:1998nq}.

\section{Dark Matter And Direct Detection}

In this section, we look to the matter content of $E_6$ GUTs for a viable candidate for the dark matter of our universe. In $E_6$, the fermions of the Standard Model come as part of three copies (one per generation) of the 27-dimensional fundamental representation. Each {\bf 27} of $E_6$ decomposes into a {\bf 16}, {\bf 10} and {\bf 1} of $SO(10)$. The fermions of the Standard Model, along with a singlet $\bar{\nu}_R$, fill the {\bf 16}, and are further decomposed into a {\bf 10} ($Q$, $\bar{u}_R$, $e^+_R$), $\bar{\bf{5}}$ ($L$, $\bar{d}_R$), and {\bf 1} ($\bar{\nu}_R$) of $SU(5)$. In this context, $\bar{\nu}_R$ may be viewed as a right handed neutrino. In addition, the {\bf 27} contains ``exotic'' fermions occupying the {\bf 10} and {\bf 1} of $E_6$. The 10 dimensional representation decomposes into a {\bf 5} ($\bar{H}$, $h_L$) and $\bar{\bf 5}$ ($H$, $\bar{h}_R$) of $SU(5)$. The state $S_L$ is a singlet under both $SO(10)$ and $SU(5)$. Thus the exotics of the {\bf 27} consist of a vector-like, color singlet, weak isodoublet with electric charges of $Q=0, \pm1$ ($H$, $\bar{H}$), a vector-like color triplet, weak isosinglet with $Q=\pm1/3$ ($h_L$, $\bar{h}_R$), and two neutral singlets ($\bar{\nu}_R$ and $S_L$). See Table~I for a decomposition of the {\bf 27} along with their gauge charges.

The matter fields in the {\bf 27} get mass terms through interactions with Higgs fields appearing in additional representations of $E_6$. In the breaking scheme of the $E_6$ $\eta$-model, at least four Higgs fields are required~\cite{Hewett:1985ss}. Two of these are $SU(2)_L$ isodoublets, corresponding to the up- and down-type Higgses of the MSSM, with hypercharges of $-1/2$ and $+1/2$. Additionally, these two fields must have $Q_{\eta}$ ($Q'$) charges of $+1\,(0)$ and $+4\,(1)$, respectively. The remaining two Higgs fields serve to give masses to the exotics in the {\bf 27}'s; these scalars must be isosinglets with $Y=0$ and have $\eta$ charges $+5$ and $+10$, giving $Q'$ charges of +1 and +2. It is the vevs of these Higgs that can serve to break $Q'$ and give mass to the leptophobic $Z'$. From the mass of the $Z'$, we expect the new Higgs vevs to be on the order of the weak scale.

Among these exotics, both $\bar{\nu}_R$ and $S_L$ are not colored, are electrically neutral, and thus are potentially viable candidates for dark matter. Furthermore, they each carry large charges of the leptophobic gauge group ($Q'=-1$), potentially providing an important role for the $Z'$ in cosmology as well as in collider physics.\footnote{Throughout, we adopt the convention of writing the $Z'$ charges as normalized in Table~I, absorbing the factor of $2\sqrt{3/5}$ into the definition of the gauge coupling constant.} We take our dark matter candidate to be the scalar superpartner of the lightest right-handed neutrino $\bar{\nu}_R$, which will be stable by the virtue of $R$-parity (this dark matter candidate has previously been considered in Ref.~\cite{Lee:2007mt}). We could also consider the superpartner of the $S_L$, which is identical for the purposes of direct detection, but is disfavored by relic abundance considerations as described in Sec.~\ref{ann}. To retain a degree of agnosticism toward this choice (or a linear combination of these choices), we will refer to the scalar dark matter particle simply as $X$.

%As the R-parity assignments of the exotic fermions (as well as their lepton and baryon number assignments) are somewhat model dependent, we can consider either the fermions $\nu^c$ or $S$, or their scalar superpartners as our dark matter candidate. With agnosticism toward this choice, we will refer to the dark matter in this model simply as $X$.

%\section{Notes}

%To accomodate the CDF $W^{\pm}$+dijets anomaly, we require a $Z'$ with a mass of $\sim$150 GeV, with substantial couplings to light quarks, but significantly smaller couplings to electrons and muons. Such a leptophobic $Z'$ naturally appears within the context of the $E_6$ Grand Unified scenario known as the $\eta$-model. 

In the $E_6$ $\eta$-model with $\delta\approx 1/3$, the light $Z'$ couples to Standard Model quarks with charges approximately given by: $Q'_{u_L}=-1/3$, $Q'_{u_R}=2/3$, $Q'_{d_L}=-1/3$, $Q'_{d_R}=-1/3$, but with little or no couplings to any Standard Model leptons. Writing the effective couplings as the product of these charges and a new gauge coupling, $g_{Z'}$, the rate observed at CDF favors a value of roughly $g_{Z'} \sim 0.7-0.8$~\cite{Buckley:2011vc}. To bring this into better compatibility with the null results of D0~\cite{D0}, however, we will adopt a somewhat smaller value of $g_{Z'} \sim 0.5$.

%The $E_6$ model in question also contains three (one per generation) exotic leptons, $n$, and right handed neutrinos, $N_e$, which have considerable charges couplings them to the leptophobic $Z'$, $Q'_{n}=Q'_{N_e}=-1$. The lightest of thest states (which may also be a mixture of $n$ and $N_e$ will be long lived, even on cosmological timescales, a will provide us with a viable dark matter candidate.

Through $Z'$ exchange, the dark matter can scatter elastically with nuclei~\cite{tchannel,Mambrini:2010dq}. The cross section for spin-independent scattering of our scalar dark matter candidate with $Q'_{X} = -1$ is given by:
%
%In the case in which the dark matter is a complex scalar, the dark matter's spin-independent elastic scattering with nuclei is given by:
%
\begin{eqnarray}
\sigma^{\rm SI}_{XN} \approx \frac{m^2_X m^2_N}{\pi (m_X+m_N)^2} \bigg[f_p Z_N+f_n (A_N-Z_N)\bigg]^2,
\end{eqnarray}
where 
\begin{eqnarray}
f_p&=&\frac{g^2_{Z'}}{m^2_{Z'}}\bigg[(Q'_{u_L}+Q'_{u_R})+\frac{1}{2}(Q'_{d_L}+Q'_{d_R})\bigg]=0, \\ 
f_n&=&\frac{g^2_{Z'}}{m^2_{Z'}}\bigg[\frac{1}{2}(Q'_{u_L}+Q'_{u_R})+ (Q'_{d_L}+Q'_{d_R})\bigg]=\frac{-g^2_{Z'}}{2 m^2_{Z'}}.\nonumber 
\end{eqnarray}
In other words, the couplings to up and down quarks cancel for the proton, leaving only neutrons with which the dark matter can scatter. With this substitution, we arrive at
\begin{eqnarray}
\sigma^{\rm SI}_{XN} &\approx& \frac{m^2_X m^2_N g^4_{Z'}}{4 \pi (m_X+m_N)^2 m^4_{Z'}} (A_N-Z_N)^2. 
\end{eqnarray}

To compare with the results of direct detection experiments, we calculate the spin-independent elastic scattering cross section of our dark matter candidate with neutrons (not to be confused with nucleons, as is often presented):
\begin{eqnarray}
 \sigma^{\rm SI}_{X-{\rm neutron}} &\approx& \frac{m^2_X m^2_n g^4_{Z'}}{4 \pi (m_X+m_n)^2 m^4_{Z'}} \\
&  \approx& 2.0 \times 10^{-39} \, {\rm cm}^2 \, \bigg(\frac{160\,{\rm GeV}}{m_{Z'}}\bigg)^4 \bigg(\frac{g_{Z'}}{0.5}\bigg)^4.\nonumber
\end{eqnarray}
%

%To compare these results to those found in the commonly addressed case of $f_p=f_n$, we write:
%%
%\begin{eqnarray}
%\sigma^{\rm SI}_{XN} (f_p=f_n) = \sigma^{\rm SI}_{X{p,n}} A^4 \bigg(\frac{m_X+m_{p,n}}{m_X+m_N}\bigg)^2,
%\end{eqnarray}
%%
%which we can now use to translate the values found in Eq.~3 to:
%%
%\begin{eqnarray}
%\sigma^{\rm SI, Eff}_{X{p,n}} ({\rm Ge}) =  6.1\times 10^{-40} \, {\rm cm}^2 \, \bigg(\frac{m_X}{7\,{\rm GeV}}\bigg)^2 \bigg(\frac{160\,{\rm GeV}}{m_{Z'}}\bigg)^4 \bigg(\frac{g_{Z'}}{0.7}\bigg)^4, \nonumber \\
%\sigma^{\rm SI, Eff}_{X{p,n}} ({\rm Na}) =  5.3\times 10^{-40} \, {\rm cm}^2 \, \bigg(\frac{m_X}{7\,{\rm GeV}}\bigg)^2 \bigg(\frac{160\,{\rm GeV}}{m_{Z'}}\bigg)^4 \bigg(\frac{g_{Z'}}{0.7}\bigg)^4,
%\end{eqnarray}
%%
%%If we scale this by the relic abundance found in Eq.~1, this leads to an effective nucleon-level cross section of $\sigma\times (\rho_{\rm local}/0.3 GeV cm^{-3})= (2.3-2.5)\times 10^{-40}$ cm$^2$ for either CoGeNT or DAMA/LIBRA, which is not at all far from the experimentally favored values.%

\begin{figure}[t]
\centering
{\includegraphics[angle=0.0,width=3.4in]{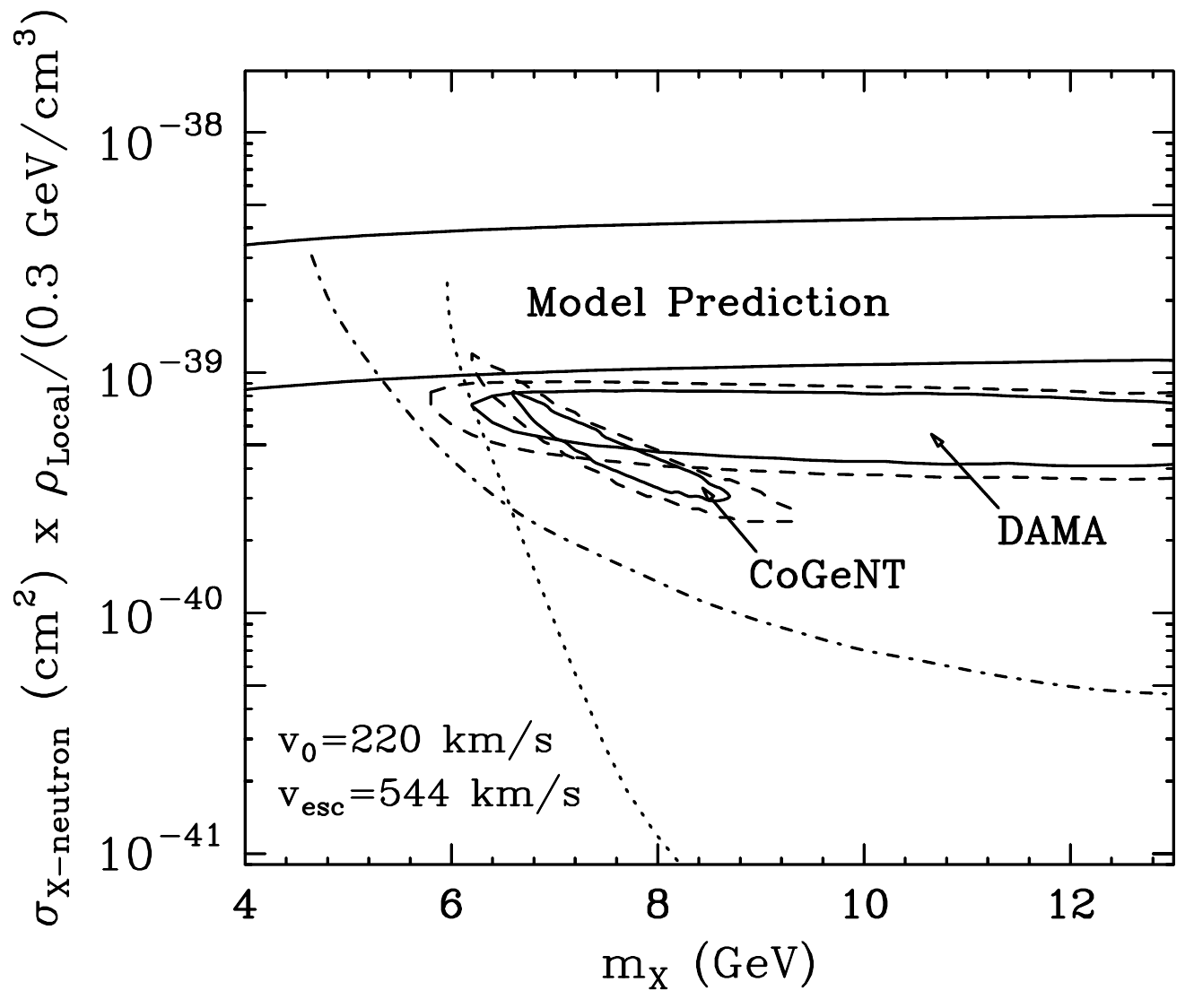}}
\caption{The dark matter's spin-independent elastic scattering cross section with neutrons (not to be confused with nucleons) predicted in the $E_6$ GUT model described in this article, compared with the regions of parameter space consistent with the signals reported by the CoGeNT~\cite{cogent} and DAMA/LIBRA~\cite{dama} collaborations (90\% and 99\% CL contours are shown~\cite{cogentdama}). To account for the uncertainty in the local dark matter density, we absorb into the cross section the scaling to this quantity, as indicated in the label of the y-axis (the width of the band corresponds to $\rho_{\rm Local}=$0.15-0.6 GeV/cm$^3$). We have adopted $g_{Z'}=0.5$ and $m_{Z'}=160$ GeV for the coupling and mass of the $Z'$, as approximately required to produce the CDF $W$+dijets excess~\cite{Buckley:2011vc} while not strongly exceeding the rate observed at D0~\cite{D0}. Also shown are the constraints presented by the CDMS~\cite{cdms} (dot-dashed) and XENON100~\cite{xenon100} (dotted) collaborations. See text for details.}
\label{direct}
\end{figure}

In Fig.~\ref{direct}, we compare this result to the cross section and mass required to explain the signals reported by the CoGeNT~\cite{cogent} and DAMA/LIBRA~\cite{dama} collaborations (see also Refs.~\cite{cogentdama}). For dark matter with a mass of approximately 6-8 GeV (and for a velocity distribution described by $v_0=220$ km/s and $v_{\rm esc}=544$ km/s), the signals reported by these two experiments can be accommodated by the $E_6$ GUT model discussed here. We also show for comparison the constraints as presented by the CDMS~\cite{cdms} (dashed) and XENON100~\cite{xenon100} (dotted) collaborations. Although the results from CDMS appear to be in some degree of tension with a dark matter interpretation of CoGeNT and DAMA/LIBRA, it is not necessarily implausible that systematic uncertainties could open a region of compatibility~\cite{Collar:2011kf}. The constraints placed on light dark matter particles by XENON100 (and other liquid xenon detectors) depend critically on the scintillation efficiency of liquid xenon at the lowest measured energies, and may well be less stringent than have been reported~\cite{Collar:2010ht}.

%Alternatively, if the dark matter is a Dirac fermion, we have the same cross section as described in Eq.~4, but with an additional enhancement of 4. which in turn leads to effective nucleon-level elastic scattering cross sections of $\sim$$10^{-39}$ cm$^2$. This can probably be reconciled with CoGeNT and DAMA if the local density is nearly an order of magnitude lower than is estimated by rotation curves.

%If instead the dark matter is a Majorana fermion, we will have to rely on spin-dependent scattering through the $Z'$.  ...

%If our dark matter candidate, $X$, is a Dirac fermion (and $m_{Z'} \gg 2 m_X > 2 m_b$), then it will annihilate through the exchange of the $Z'$ to Standard Model quarks with a cross section given by:
%%
%\begin{eqnarray}
%\sigma v &\approx& \frac{m^2_X}{2 \pi} \sum_q \frac{3 g^4_{Z'} Q'^2_q}{m^4_{Z'}} (1-m^2_q/m^2_X)^{1/2} \bigg[2 + \frac{m_q^2}{m^2_X}+\mathcal{O}(v^2) \bigg], \nonumber \\
%&\approx& 1.3 \times 10^{-25} \, {\rm cm}^3/{\rm s} \, \bigg(\frac{m_X}{7\,{\rm GeV}}\bigg)^2 \bigg(\frac{150\,{\rm GeV}}{m_{Z'}}\bigg)^4 \bigg(\frac{g_{Z'}}{0.75}\bigg)^4,
%\end{eqnarray}
%%
%which is a factor of $\sim$4  larger than that required to produce the observed dark matter abundance (the $X$ will make up only $\sim$20\% of the dark matter).%

\section{Dark Matter Relic Abundance and Indirect Detection}
\label{ann}

In this section, we turn our attention to the annihilation of the dark matter candidate appearing in the $\eta$-model of $E_6$. First, we consider annihilations through the $s$-channel exchange of the $Z'$, which leads to a cross section given by:
\begin{eqnarray}
\sigma v_{XX\rightarrow q\bar{q}} \approx \frac{m^2_X g^4_{Z'} v^2}{4 \pi m^4_{Z'}} \sum_q Q'^2_q \sqrt{1\!-\!\frac{m^2_q}{m^2_X}}\bigg[1 \!+ \!\frac{m_q^2}{2 m^2_X}\bigg],
\end{eqnarray}
where $v$ is the relative velocity between two dark matter particles and the sum is performed over all favors of left and right-handed quarks lighter than $m_X$. Note that the annihilation cross section is velocity suppressed (i.e.,~$\sigma v \simeq b v^2$). To determine the thermal relic abundance of dark matter, we consider the thermally weighted cross section at the temperature of freeze-out:
\begin{eqnarray}
\langle \sigma v \rangle  &\approx& \frac{3b}{x_{\rm FO}} \approx 1.2 \times 10^{-27} \, {\rm cm}^3/{\rm s} \\
&\times& \bigg(\frac{m_X}{7\,{\rm GeV}}\bigg)^2 \bigg(\frac{160\,{\rm GeV}}{m_{Z'}}\bigg)^4 \bigg(\frac{g_{Z'}}{0.5}\bigg)^4 \bigg(\frac{20}{x_{\rm FO}}\bigg), \nonumber
\end{eqnarray}
where $x_{\rm FO}$ is the ratio of $m_X$ to the freeze-out temperature. The dark matter's annihilation cross section is related to its thermal relic abundance according to
\begin{equation}
\Omega_{X} h^2 \approx 0.1 \times \bigg(\frac{3\times 10^{-26} \, {\rm cm}^3/{\rm s}}{\langle \sigma v \rangle}\bigg).
\end{equation}
Annihilations of dark matter through the $Z'$ thus provide only a small fraction ($\sim$5\%) of the cross section required to yield our universe's observed dark matter density. Therefore processes other than $Z'$ exchange must contribute to the dark matter's annihilation in the early universe.

%We could instead consider the case in which the dark matter consists of the superpartners of the $\nu^c$ and/or $S$, a complex scalar. In this case, the annihilation cross section is velocity suppressed ({\it ie.} $\sigma v \simeq b v^2$):
%%
%\begin{eqnarray}
%\sigma v &\approx& \frac{m^2_X v^2}{6 \pi} \sum_q \frac{3 g^4_{Z'} Q'^2_q}{m^4_{Z'}} (1-m^2_q/m^2_X)^{1/2}\bigg[1 + \frac{m_q^2}{2 m^2_X}\bigg],
%\end{eqnarray}
%which after thermally averaging at the time of relic freeze-out, $<\sigma v> \approx 3b/x_{\rm FO}$, we find
%\begin{eqnarray}
%<\sigma v> &\approx& 3.3 \times 10^{-27} \, {\rm cm}^3/{\rm s} \\
%&\times& \bigg(\frac{m_X}{7\,{\rm GeV}}\bigg)^2 \bigg(\frac{150\,{\rm GeV}}{m_{Z'}}\bigg)^4 \bigg(\frac{g_{Z'}}{0.75}\bigg)^2 \bigg(\frac{20}{x_{\rm FO}}\bigg), \nonumber
%\end{eqnarray}
%%
%where $x_{\rm FO}$ is the ratio of  $m_X$ to the freeze-out temperature.

%Alternatively, we could consider dark matter annihilations through diagrams other than $Z'$ exchange. 

The most general renormalizable superpotential describing the interactions within the {\bf 27} that is consistent with the symmetries of the Standard Model is given by:
\begin{eqnarray}
W_{\rm Int}&=&\lambda_1 \bar{H} Q \bar{u}_R + \lambda_2 H Q \bar{d}_R + \lambda_3 H L e^+_R +  \lambda_4 \bar{H} H S_L \nonumber \\
&+&   \lambda_5 h_L \bar{h}_R S_L+ \lambda_6 h_L \bar{u}_R e^+_R +  \lambda_7 L \bar{h}_R Q +  \lambda_8 \bar{\nu}_R h_L \bar{d}_R \nonumber \\
&+& \lambda_9 h_L QQ  +  \lambda_{10} \bar{h}_R \bar{u}_R \bar{d}_R + \lambda_{11} \bar{H} L \bar{\nu}_R.  
\end{eqnarray}
Depending on the lepton and baryon number assignments of the exotic particle content of the {\bf 27}, some of these terms will need to be suppressed in order to maintain approximate low-scale lepton and baryon number conservation.

Among these 11 terms, there are no interactions which directly connect the $S_L$ (or its superpartner) to the fields of the Standard Model. For this reason, we will focus our attention on the case in which the dark matter is the superpartner of $\bar{\nu}_R$. The interactions described by the terms $\lambda_8 \bar{\nu}_R h \bar{d}_R$ and $\lambda_{11} \bar{H} L \bar{\nu}_R$ allow the dark matter to annihilate into down-type quarks through the $t$-channel exchange of a $h_L$ or to leptons through the $t$-channel exchange of a $\bar{E}^{-}$ or $\bar{N}^0$. As null results from the LHC force states with QCD color such as $h_L$ to be quite heavy (above several hundred GeV), we focus on the case of annihilation to charged and neutral leptons through a $\bar{E}^{-}$ or $\bar{N}^0$, respectively. Assuming for simplicity that $m_{\bar{E}^{-}}$, $m_{\bar{N}^0}$ and $\lambda_{11}$ have similar values for each generation (in each {\bf 27}),  the dark matter's annihilation cross section to charged leptons and neutrinos is approximately given by~\cite{tchannel}:
\begin{eqnarray}
\sigma v \approx \frac{\lambda^4_{11} \, m^2_{\tau}}{16 \pi\, m^4_{E^-}} + \frac{v^2 \, \lambda^4_{11}\, m^2_X}{48 \pi}\sum_{l=e,\mu,\tau} \bigg[\frac{1}{m^4_{E^{-}}}+ \frac{1}{m^4_{N^{0}}} \bigg],
\label{tch}
\end{eqnarray}
where the first term proceeds to $\tau^+ \tau^-$ while the terms in the sum proceeds to a mixture of charged leptons and neutrinos. In the low-velocity limit, the annihilation cross section to $\tau^+ \tau^-$ given by:
\begin{eqnarray}
\sigma v \approx 4.7\times 10^{-27} \, {\rm cm}^3/{\rm s} \,\times \, \bigg(\frac{\lambda_{11}}{1}\bigg)^4 \,\bigg(\frac{110 \, {\rm GeV}}{m_{E^-}}\bigg)^4. 
\end{eqnarray}
Thus for the purposes of indirect detection, the dark matter will annihilate primarily to $\tau^+ \tau^-$, possibly with an order $\sim$10\% contribution to $b$-quarks through $t$-channel $h_L$ exchange. We note that in order for this annihilation process to proceed, either (but not both of) $\bar{\nu}_R$ or $\bar{H}$ must carry lepton number.

After thermally averaging at the temperature of freeze-out, the $v^2$ terms of the cross section given in Eq.~\ref{tch} becomes:
\begin{eqnarray}
\langle \sigma v \rangle &\approx& 2.3 \times 10^{-26} \, {\rm cm}^3/{\rm s} \\
&\times& \bigg(\frac{\lambda_{11}}{1}\bigg)^4 \bigg(\frac{110 \, {\rm GeV}}{m_{E^-,N^0}}\bigg)^4  \bigg(\frac{m_X}{7 \, {\rm GeV}}\bigg)^2 \bigg(\frac{20}{x_{\rm FO}}\bigg). \nonumber
\end{eqnarray}
Combining this with the results from Eqs.~7 and~11, we find a total value which is very similar to that required to thermally generate the observed density of dark matter.

\begin{figure}[t]
\centering
{\includegraphics[angle=0.0,width=3.4in]{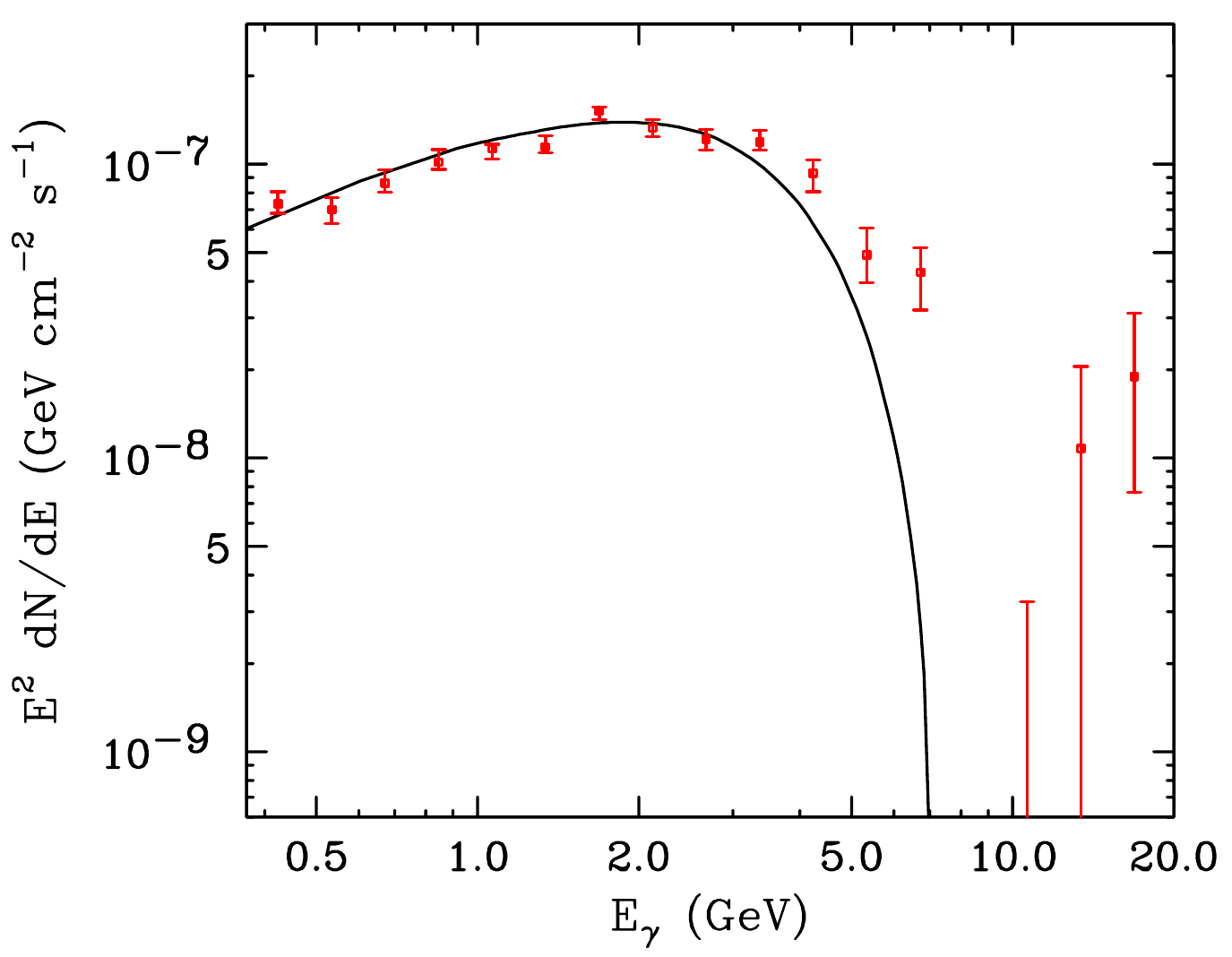}}
\caption{The spectrum of gamma rays from the inner degree around the Galactic Center produced from annihilations of the right-handed sneutrino dark matter candidate present in the $E_6$ model described in this paper, compared with the excess emission observed by the Fermi Gamma Ray Space Telescope (after subtracting known astrophysical backgrounds, including an extrapolation of the emission from the Milky Way's supermassive black hole)~\cite{lisa}. Here, we have taken $m_X=8$ GeV, $m_{\bar{E}}=110$ GeV, $m_{h_L}=750$ GeV, $\lambda_{11}=1$, $\lambda_{8}=1$, and $\rho_{X}=0.3$ GeV/cm$^3 \times (8.5\, {\rm kpc}/R)^{1.3}$, where $R$ is the distance from the Galactic Center.}
\label{indirect}
\end{figure}

The leptophilic annihilations of the dark matter is of particular interest within the context of gamma ray observations of the Galactic Center. In Fig.~\ref{indirect}, we compare the spectrum of gamma rays from dark matter annihilations in the Galactic Center predicted in our model to the excess emission observed by the Fermi Gamma Ray Space Telescope~\cite{lisa} (assuming an extrapolation of the spectrum from the Milky Way's supermassive black hole; see also Ref.~\cite{nofermi}). Clearly, very good agreement is found. Dark matter in the mass range described here and which annihilates primarily to charged leptons has also been previously shown to be able to produce the observed synchrotron emission known as the WMAP Haze~\cite{hazelight} (see also Ref.~\cite{haze}). 
Lastly we note the constraints from searches for GeV-scale neutrinos from the Sun as conducted by Super-Kamiokande~\cite{superK} are safely evaded in this model by virtue of the dark matter's suppressed couplings to protons.

\section{Summary and Conclusions}

In summary, motivated by the $W$+dijets excess recently reported by the CDF collboration, we have considered Grand Unified Models based on the $E_6$ group. Such models (in particular, the so-called $\eta$-model of $E_6$), can contain a leptophobic $Z'$ boson capable of producing the observed excess. Within the fundamental representation of $E_6$ is an attractive dark matter candidate in the form of a right-handed sneutrino. From the sizable couplings of this dark matter candidate to the $Z'$, a large elastic scattering cross section with neutrons is predicted, consistent with the direct detection signals reported by the CoGeNT and DAMA/LIBRA collaborations. Furthermore, to avoid being overproduced in the early universe, the dark matter must annihilate to leptons through the exchange of other particles contained within the fundamental representation of $E_6$, leading to a gamma ray spectrum and flux consistent with that observed from the Galactic Center by the Fermi Gamma Ray Space Telescope.

\smallskip

%We would like to thank ... for valuable discussions. 

MRB and DH are supported by the US Department of Energy. DH is also supported by NASA grant NAG5-10842. JLR is supported in part by the US Department of Energy under Grant No.\ DE-FG02-90ER40560.

\end{document}